\documentclass[useAMS,usenatbib]{esann}
\usepackage[dvips]{graphicx}
\usepackage[latin1]{inputenc}
\usepackage{amssymb,amsmath,array}
\usepackage{macros}
\usepackage{color}
\usepackage{adjustbox}
\usepackage{txfonts}
\usepackage{siunitx}
\usepackage{subfigure}
\usepackage{wrapfig}
\begin{document}
\title{Stellar formation rates in galaxies using Machine Learning models}

\author{Michele Delli Veneri$^1$, Stefano Cavuoti$^{1,2,3}$, Massimo Brescia$^2$, Giuseppe Riccio$^2$ \and 
and Giuseppe Longo$^{1,3}$
%
%
\vspace{.3cm}\\
%
1- Universit\'a degli Studi Federico II - Dipartimento di Fisica  ``E. Pancini'',\\
via Cintia 6, I-80135 Napoli, Italia\\
\vspace{.1cm}\\
2- INAF - Osservatorio Astronomico di Capodimonte,\\
via Moiariello 16, I-80131 Napoli, Italia\\
\vspace{.1cm}\\
3- INFN - Napoli Unit, via Cintia 6, I-80135 Napoli, Italia 
}

\maketitle

\begin{abstract}
Global Stellar Formation Rates or SFRs are crucial to constrain theories of galaxy formation and evolution. 
SFR's are usually estimated via spectroscopic observations which require too much previous telescope time and therefore 
cannot match the needs of modern precision cosmology. We therefore propose a novel method to estimate SFRs for
large samples of galaxies using a variety of supervised ML models.
\end{abstract}
\section{Introduction}
One of the main parameters used to understand galaxy formation and evolution is the global Stellar Formation Rate (SFR) \cite{madau} which measures the luminosity-weighted average across local variations in star formation history and physical conditions within a given galaxy. The identification of proper SFR indicators and their calibration has been at the center of intensive debate for the last thirty years and many problems must yet be solved. However, in a broad oversimplification, global SFR indicators are measures of luminosity, either monochromatic or integrated over some wavelength range chosen in order to be sensitive to the short-lived massive stars. There is a large variety of such indicators spanning from the UV/optical/near-IR range ($\sim$ 0.1-5 $\mu m$) which probe the direct stellar light emerging from young stars,  to the mid/far-IR ( $\sim$ 5-1000 $\mu m$) probing the stellar light reprocessed by dust. Other indicators rely on the ionizing photon rate, as traced by the gas ionized by massive stars \cite{calzetti2004,hong2011};  hydrogen recombination lines, forbidden metal lines, and in the millimeter range, the free-free (Bremsstrahlung) emission. Finally, other indicators are, at least in principle, derived in the X ray domain, from X-ray binaries, massive stars, and supernovae; the non-thermal synchrotron emission produced by accelerated charged particles, which, following the suggestions in \cite{condon1992}, can be calibrated as an SFR indicator.

\begin{figure}\centering
\includegraphics[width=.8\textwidth]{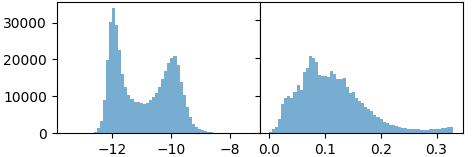} 
\caption{\textit{Left panel:} distribution of AVG; \textit{right panel:} Spectroscopic redshift distribution of the training sample.} \label{fig:specz}
\end{figure}

It is clear, however, that the correct evaluation of SFR from optical/FIR broad band data is a highly non trivial task due to the complex  and still largely poorly understood correlation existing between the SFR and the broad band photometric properties integrated over the whole galaxy. 
From the mathematical point of view, the estimation of SFR is a typical regression problem and, in a recent work \cite{Stensbo} it has been shown how machine learning methods can be effectively used to estimate stellar formation rates for objects in large panchromatic sky surveys.
In this work we make use of the SFR's derived in \cite{Brinchmann} for a subsample of galaxies in the SDSS-DR7. While referring to the original paper for a detailed description, we shall just remind that Brinchman \cite{Brinchmann}  used the Charlot \& Longhetti model \cite{charlot2001} and that the method to obtain SFR estimates for other classes of galaxies (AGN, Composite, low S/N SF, low S/N LINER and Unclassifiable) is nearly identical to that described in \cite{Brinchmann}.

\section{Data}\label{SEC:data}
In order to compare the results of our work with those obtained by \cite{Stensbo} we used the Data Release 7 of the Sloan Digital Sky Survey (hereafter SDSS-DR7) \cite{DR7}. 
In fact this data release was used in \cite{Brinchmann} to derive reliable star formation rates for a subsample of $\sim 10^6$ galaxies with good spectroscopy. 
To create our Knowledge Base (or KB), we extracted from the \textit{SpecPhoto} view of the SDSS-DR7 all objects whose PlateID, FiberID and MJD matched the objects in the the Brinchmann et al. catalogue \cite{Brinchmann}. From the former we extracted the \textit{psfMag}, \textit{fiberMag}, \textit{petroMag}, \textit{modelMag}, \textit{expMag} and \textit{deVMag} magnitudes in the u, g, r, i, and z bands and the \textit{spectroscopic redshift}; from the latter, the targets i.e. the average total star formation rates (\textit{AVG})\footnote{ "Magnitudes" $m$ are the conventional astronomical equivalent of the fluxes $f_{\lambda}$. More precisely $m=-2.5 Log f_{\lambda} +q_0$, where $q_0$ is a zero point. }. Then the following constrains were applied:
\begin{enumerate}
\itemsep0em 
\item for SFRs we required a successful estimation (\textit{flag} = 0 in \cite{Brinchmann});
\item for redshifts we required a successful spectroscopic estimation (\textit{}{zWarning} = 0 in \cite{DR7})
\end{enumerate}

A sample of 572,285 galaxies satisfied those criteria. Then we eliminated from the sample all objects with a "\textit{Null}" feature and truncated all magnitude distributions to physically meaningful values, obtaining a KB of 570,908 galaxies. For each magnitude type, we produced the \textit{u - g}, \textit{g - r}, \textit{r - i}, and \textit{i - z} colors, reaching a total of 54 photometric features for our KB plus the AVG as a target. The distribution of spectroscopic redshifts and AVGs for the KB is shown in Fig. \ref{fig:specz} (right and left panel, respectively)	.
To study the relationship between redshift and star formation rate, we cross-matched the sample of 570,908 galaxies with our catalogue of photometric redshifts for the SDSS-DR9 galaxies \cite{Brescia2014}, retaining 436,490 galaxies out of the 570,908. Thus integrating the features list with: the spectroscopic redshift \textit{zSpec} and the relative photometric redshift estimations \textit{zPhot} derived from \cite{Brescia2014}, we reached a 56-dimensional parameter space (hereafter DR7-KB).
Furthermore we created a second database based on the first one (DR9-KB) using the more accurate photometric data from the SDSS-DR9 \cite{dr9}.

\begin{figure}
\centering
\includegraphics[width=1.0\textwidth]{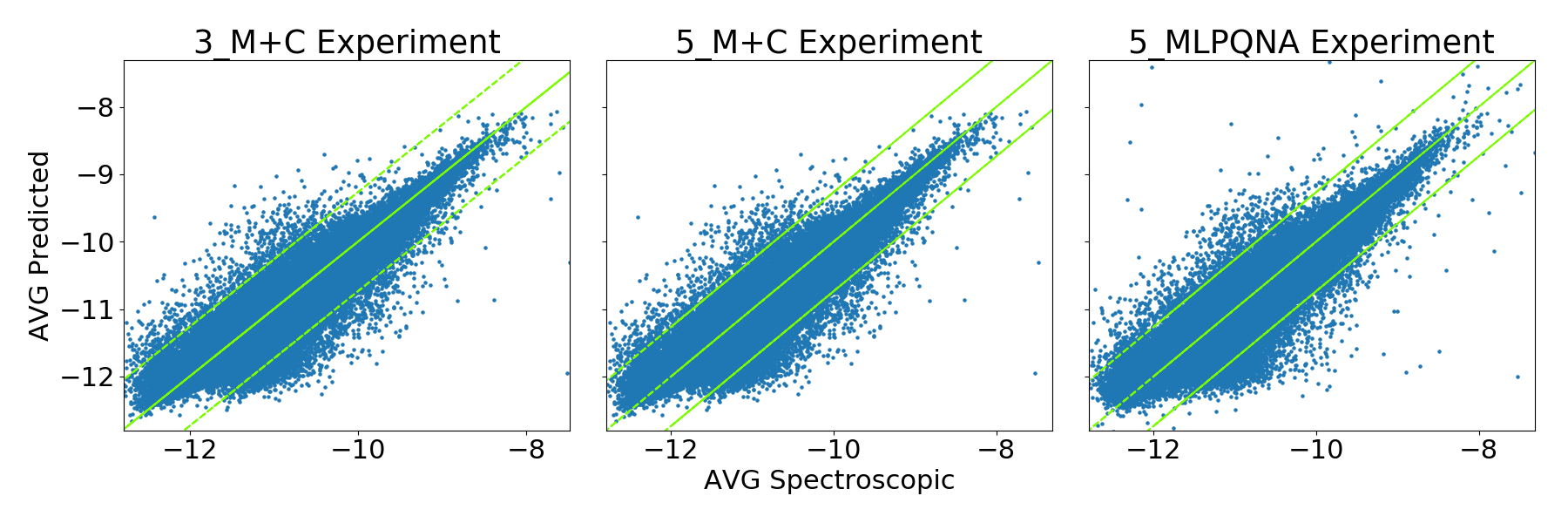}
\caption{Spectroscopic AVG VS Predicted AVG.The central line gives the bisector, while the other two define the $3\sigma$ regions. Objects outside these \textbf{green lines} must be considered as "outliers".}\label{fig:figavg}
\end{figure}

\section{Algorithms}\label{SEC:Algorithms}

In the present work we make use of two Machine Learning (ML) methods, respectively Random Forest (RF) \cite{Breiman20015} and Multi Layer Perceptron trained by the Quasi Newton Algorithm (MLPQNA) \cite{Cavuoti201545}.
In order to optimize their performances, we apply the k-fold cross validation  \cite{Kohavi95astudy} and two feature selection methods. The latter are respectively, the forward selection \cite{Mao2004629} based on the feature importance embedded in the RF and PHiLAB (Parameter Handling including Lasso Analysis as Benchmark). PHiLAB (Brescia et al. in prep.) is a method designed to solve the known problem of the all-relevant feature selection, i.e. the identification of the exact parameter space that could be considered as relevant in the context of a problem solution analysis. By performing an exhaustive selection of all relevant features, it enables a deeper understanding of the physical phenomena  governing the specific problem (in this case the star forming rate prediction). The importance of any feature of the given parameter space is evaluated as the accuracy loss induced by a random perturbation of the original feature among  the available data entries. For each feature the relative measure of  significance is then compared to the statistical score obtained by evaluating the regression estimation, by means of the known LASSO statical method \cite{hara2016}, based on the L1-norm regularization. 
The combination of the features supporting the optimal solution of  Lasso regression with those out-coming from the random noising mechanism, is able to furnish a robust and complete set of features composing the best parameter space for the current problem \cite{hara2016}.

\section{Experiments}\label{sec:exp}
We performed a series of experiments divided in five groups (as listed in Table \ref{tab:tab1}).
In the first group, we used the full DR7-KB (54 or 55 features depending on the experiment) and the k-fold cross validation in order to compare our results with the ones obtained in \cite{Stensbo}; in the second, we used the same KB but, instead of the k-fold cross validation, we used the traditional train-validation-test approach \cite{bishop2007} in order to evaluate the cross validation effect on the performances of our models. In the third group, we replicated the second group experiments by replacing the full set of features with the most informative ones selected by the PHiLAB method.
In the fourth group we replicated the second group experiments on the DR9-KB, in order to identify the best model to be used to produce the final SFR catalogue as well to compare the effects of different photometry. In the fifth group, we replicated the third group of experiments on the DR9 KB, using only the most informative features selected by the PHiLAB  method.
The experiment ID (see Table \ref{tab:tab1}) is composed of a number and an acronym; the number specifies the group while the following acronym indicates the set of features:
\begin{itemize}
\itemsep0em 

\item M+C: \textit{Magnitudes + Colors}. We used as input features the combination of magnitudes and colors described in section  \ref{SEC:data}  and as a target the \textit{AVG};
\item ZSPEC: \textit{Spectroscopic Redshifts}. We added to the M+C parameter space the spectroscopic redshift  (\textit{zSpec}) to check if a correlation between SFR and  spectroscopic redshifts exists;
\item ZPHOT: \textit{Photometric Redshifts}. We added to the M+C parameter space the photometric redshifts estimates (\textit{zPhot}) to check if a correlation  between SFR and photometric redshifts exists;
\item MLPQNA. We used the same KB used for the ZSPEC experiment in the third group and for the M+C experiment in the fifth group but, instead of the RF, we used the MLPQNA described in \cite{Brescia2013}.
\end{itemize}

\section{Results, Conclusions and Future Works}\label{sec:result}
\begin{figure}
\centering
\includegraphics[width=1.0\textwidth, trim={0 0 0 0cm},clip]{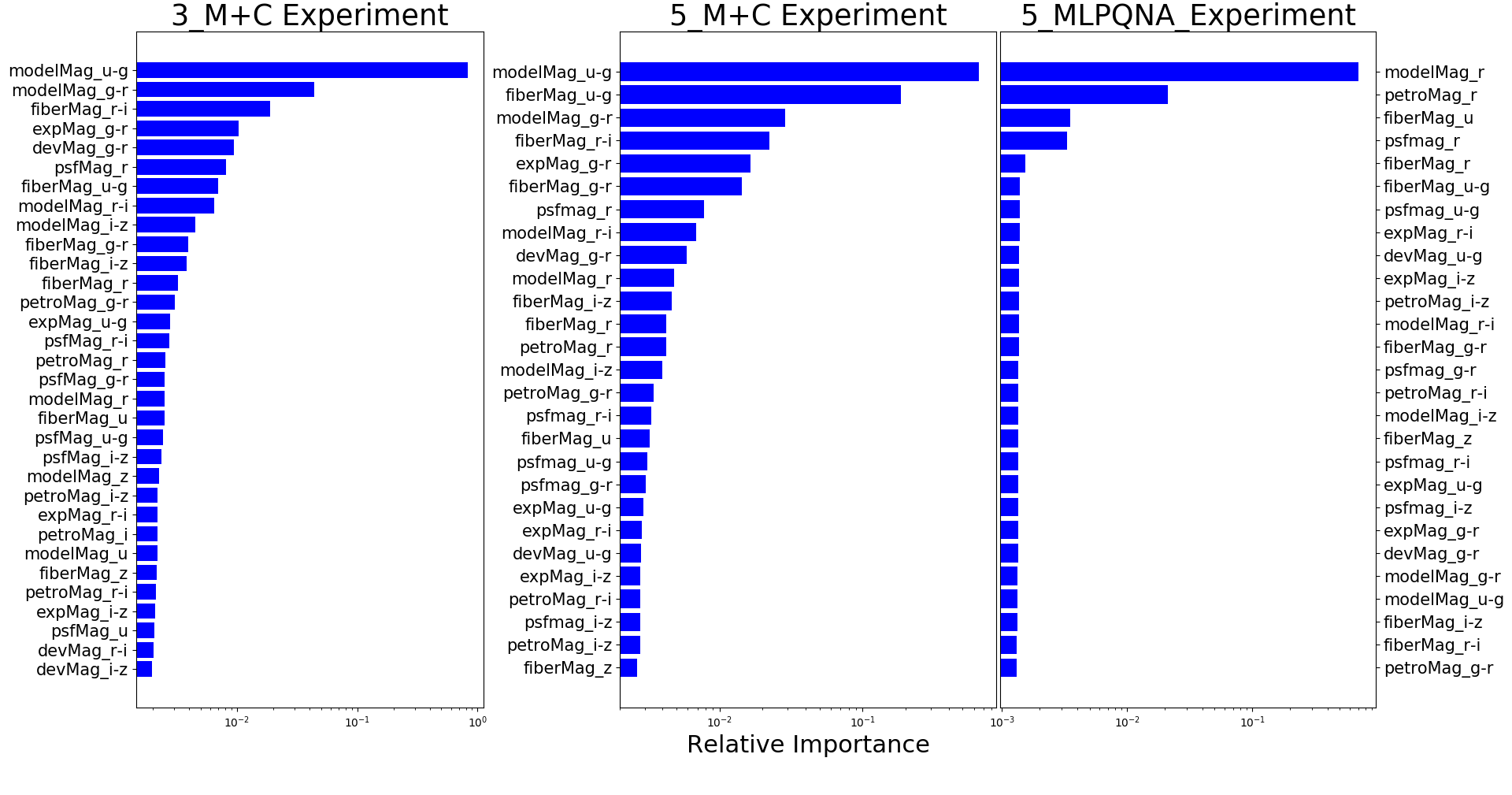}
\caption{Feature Importance Histogram}\label{fig:featimp}
\end{figure}

The statistical results of the experiments described above are summarized in Table \ref{tab:tab1}. 
In Fig. \ref{fig:figavg} we plot the spectroscopic AVG vs our output for three experiments: 3\_M+C (best experiment on DR7-KB), 5\_M+C (same as previous experiment but applied to the DR9-KB) and, finally, 5\_MLPQNA (best experiment on DR9-KB).
In Fig. \ref{fig:featimp} we show the feature importance for those three experiments.
In particular, by focusing the attention on RMSE and $\eta_{\%}$ (percentage of outliers having estimates outside the ($>3\sigma$) boundaries) quantities, group 1 experiments show that the k-fold cross validation technique does not improve the performance  on such KB.
From group 2 and 3 it appears evident that the information about photometric redshifts does not improve the SFR estimation performance. In fact, this feature is always relegated at the bottom of the feature importance ranking list.  This could be explained by keeping in mind that, due to obvious selection effects, in this work only low redshift galaxies ($z_{spec} < 0.33$) are been evaluated (see Fig. \ref{fig:specz})  and thus the photometric redshift does not carry enough information to boost the performance. On the other hand, the introduction of the spectroscopic redshift in our KB, significantly improved the performance of our models but, due to the small percentage of the SDSS-DR9 that has measured spectroscopic redshifts which are not already contained in the KB, this information is of little or no practical utility in producing a wider catalogue of SFRs. From Table \ref{tab:tab1}, by looking at group four and five, it is evident that our performance on the DR9-KB is worst than that on the DR7-KB; this can be traced to the changes in photometry between the two data releases (changes reflected by the differences in the feature ranking as shown in Fig. \ref{fig:featimp}). Regardless of the general improvement of the DR9 photometry over the DR7, the AVG estimation depends from the DR7 photometry \cite{Brinchmann} and thus it carries over all its intrinsic errors later corrected in the DR9. At this stage of our work, we believe that the loss in performance is acceptable ($\sim 6\%$). 
At the moment we are performing also other experiments using different ML models and different types of SFR estimators (i.e. \textit{specific SFR} \cite{Brinchmann}). This will enable the possibility of a direct comparison between our experiments and those described in \cite{Stensbo}. We wish however to emphasize that these results are the best estimates of SFRs obtained so far with ML based methods and that they can therefore provide a viable alternative to the spectroscopic approach.

\begin{table*}
\tiny
    \setlength{\tabcolsep}{4pt} 

    \centering
	\caption{Results of the experiments; in column 1 we have the ID, in column 2 the size of the parameter space, in column 3 the used model, in column 4 the KB, in column 5 the root mean squared error, in column 6 the median, in column 7 the standard deviation and finally in column 8 the percentage ($\eta_{\%}$) of outliers having estimates outside the ($>3\sigma$) boundaries. }
    \begin{adjustbox}{width=.985\textwidth}
	\begin{tabular}{|l | c | c | c | c | c | c | c|}
		\hline
		ID & feat. & Model & KB & RMSE & Median & $\sigma$ & $\eta_{\%}$ \\
		\hline
        1\_M+C & 54 & RF & DR7 & $0.247$ & $-0.022$ & $0.247$  & $2.08$\\
        \hline
        1\_ZSPEC & 55 & RF & DR7 & $0.229$& $-0.019 $ & $0.229$ & $2.26$  \\
        \hline
        1\_ZPHOT & 55 & RF & DR7 & $0.247$  & $-0.022$ & $0.247$& $2.06$\\
        \hline
         2\_M+C & 54 & RF & DR7 & $0.245 $ & $-0.022$ & $0.245$  & $2.07$\\
         \hline
        2\_ZPHOT & 55 & RF & DR7 & $0.245 $  & $-0.022$ & $0.245$& $2.05$\\
        \hline
        3\_M+C & 32 & RF & DR7 & $0.245 $ & $-0.022$ &$0.245$ & $2.12$\\
        \hline
        3\_ZPHOT & 33 & RF & DR7 & $0.245 $ & $-0.022$ &$0.245$ & $2.10$\\
        \hline
        3\_ZSPEC & 33 & RF & DR7 & $0.227 $ & $-0.020$ &$0.227$ & $2.32$\\
        \hline
        3\_MLPQNA & 33 & MLPQNA & DR7 & $0.218 $ & $-0.015$ &$0.218$ & $2.26$\\
        \hline
        4\_M+C & 54 & RF & DR9 & $0.261 $ & $-0.020$ &$0.261$ & $1.95$\\
        \hline
        4\_ZPHOT & 55 & RF & DR9 & $0.262 $ & $-0.019$ &$0.262$ & $1.89$\\
        \hline
        5\_M+C & 27& RF& DR9& $0.261$& $-0.019$& $0.261$& $1.94$\\
        \hline
        5\_ZPHOT & 28& RF& DR9& $0.262 $ & $-0.019$ &$0.262$ & $1.93$\\
        \hline 
        5\_MLPQNA & 27 & MLPQNA & DR9 & $0.254$ & $-0.015$ & $ 0.254$ & $2.05$\\
        \hline
	\end{tabular} \label{tab:tab1}
    \end{adjustbox}
\end{table*}

\begin{footnotesize}

\bibliographystyle{unsrt}
\bibliography{references}

\end{footnotesize}


\end{document}